\documentclass[aps,twocolumn,showpacs]{revtex4}
\usepackage{graphicx}
\begin{document}

\title{Field-Effect Transistors on Tetracene Single Crystals}
\author{R.W.I. de Boer, T.M. Klapwijk, and A.F. Morpurgo}
\affiliation{Department of Nanoscience, Faculty of Applied
Sciences, Delft University of Technology, Lorentzweg 1, 2628 CJ
Delft, the Netherlands}
\date{\today}

\begin{abstract}
We report on the fabrication and electrical characterization of
field-effect transistors at the surface of tetracene single
crystals. We find that the mobility of these transistors reaches
the room-temperature value of $0.4 \ cm^2/Vs$. The non-monotonous
temperature dependence of the mobility, its weak gate voltage
dependence, as well as the sharpness of the subthreshold slope
confirm the high quality of single-crystal devices. This is due to
the fabrication process that does not substantially affect the
crystal quality.
\end{abstract}


\maketitle

Common strategies for the fabrication of organic field-effect
transistors (FETs) are based on thin-film technology
\cite{Crone01,Deleeuw00}. This choice is motivated by the existing
deposition techniques for organic thin films that facilitate
device fabrication. At the same time, thin films usually contain a
considerable amount of structural imperfections, which affect
negatively the transistor performance \cite{Campbell01}.

For small organic molecules, crystalline films can be used to
reduce the amount of structural defects. It has been found,
however, that also the performance of transistors based on these
films are affected by structural imperfections, even when only one
crystalline grain is present between source and drain
\cite{Schoonveld98}. This is due to disorder present in the first
few molecular layers, in contact with the substrate, which
constitute the device active region \cite{Dimitrakopoulos02}. For
this reason, improving the quality of organic thin-film
transistors (TFTs) requires highly ordered molecular films, in
which the order extend up to the interface with the substrate.

An alternative route to the production of high-quality organic
FETs is to fabricate devices on the surface of a free-standing
single crystal of organic molecules. If a fabrication process that
preserves the quality of the crystals \cite{Karl85,Grosso} can be
developed, the resulting single-crystal FETs should perform better
than their TFT counterpart. Whereas considerable amount of work is
currently aiming at improving the quality of organic TFTs
\cite{Schoonveld98,Klauk02,Nelson98,Deleeuw00,Dimitrakopoulos02},
the investigation of single-crystal organic FETs has received only
limited attention
\cite{Podzorov03,Podzorov03a,Butko03,Takeya03,Ichikawa02}.

In this paper we discuss the fabrication and electrical
characterization of field-effect transistors at the surface of
tetracene single crystals. The fabrication process is based on
adhesion of pre-grown, free-standing crystals to a thermally
oxidized Si wafer on which source and drain electrodes are
deposited in advance. As we will show, this process preserves the
quality of the starting crystals. From the electrical evaluation
of a large number of devices we find that the mobility of the
charge carriers (holes) in our single-crystal FETs is reproducibly
high, reaching $0.4 \ cm^2/Vs$ in the best device. In addition,
the observed temperature and gate voltage dependence of the
mobility as well as the subthreshold slope indicate that the
performance of single-crystal devices compare well to the best
existing organic thin film transistors \cite{Campbell01}.

\begin{figure}[b]
\centering
\includegraphics[width=7.5cm]{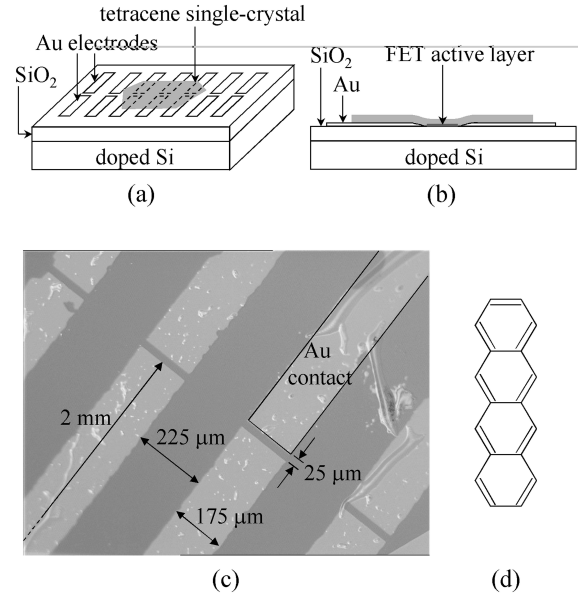}
\caption{Schematic representation (a) and side view (b) of a
tetracene single-crystal FET. (c) Optical microscope image of a
tetracene single-crystal FET. In this device, the semi-transparent
tetracene single crystal extends over several pairs of electrodes,
which are clearly visible under it. In most cases smaller crystals
have been used which extend over only one or two pair of contacts.
These different configurations allow us to study transistors with
different $W/L$ ratios on the same crystal. (d) Molecular
structure of the tetracene molecule. \label{FETpicture}}
\end{figure}

The FET fabrication involves two main steps: the growth of single
crystals and the preparation of a substrate on which the crystal
is subsequently placed \endnote{A similar technique has been used
for crystals of different molecules, see \cite{Takeya03} and
\cite{Ichikawa02}.}. Tetracene single crystals are grown by means
of physical vapor deposition in a temperature gradient in the
presence of a stream of Argon gas. The set-up used for the crystal
growth is similar to that described in ref. \cite{Laudise98}. The
source material is 98\% pure tetracene purchased from
Sigma-Aldrich. Crystals grown from as-purchased tetracene are used
as source material for a subsequent re-growth process, which
results in crystals of increased chemical purity. Tetracene
crystals grown using physical vapor deposition are platelets. For
the devices described in this paper we select thin ($\sim 1 \ \mu
m$ thick) single crystals obtained by stopping the second
re-growth process at an early stage.

\begin{figure}[t]
\centering
\includegraphics[width=8.5cm]{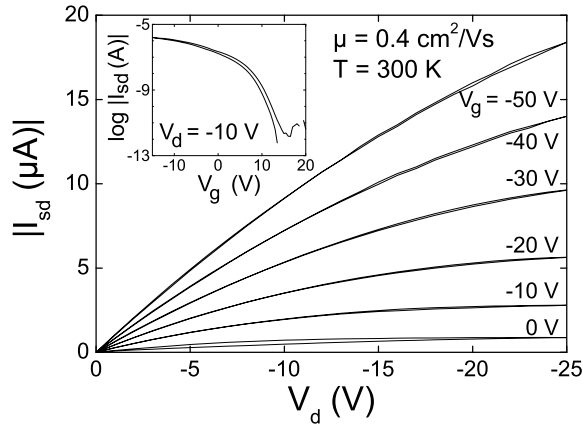}
\caption{Source-drain current $I_{sd}$ versus drain voltage $V_d$
measured at different values of $V_g$. The inset shows the
dependence of $\log (I_{sd})$ on $V_g$ at fixed $V_d$, for a
different device, which has a mobility $\mu = 0.05 \ cm^2/Vs$ and
a threshold voltage $V_t \simeq 0.3 \ V$. From this plot we
calculate the subthreshold slope to be $1.6 \ V/ \mathrm{decade}$.
\label{FETcharacteristics}}
\end{figure}

The substrates used for the FET fabrication are highly doped
(n-type or p-type) Si wafers, covered with a layer of $200 \ nm$
thermally grown SiO$_2$. The conducting Si wafer serves as gate
electrode, with the SiO$_2$ layer acting as gate insulator. Gold
contacts are deposited on top of the SiO$_2$ by means of e-beam
evaporation through a shadow mask (see fig. \ref{FETpicture} for
the precise geometry and dimensions). The dimensions of the Au
contacts as well as the tetracene crystal size determine the
transistor channel length $L$ and channel width $W$. This allows
us to study transistors with different $W/L$ ratios. Prior to
placing the tetracene single crystals on top of the substrate, the
SiO$_2$ surface is cleaned by Reactive Ion Etching (RIE) in an
oxygen plasma. We found that this cleaning step is crucial for
reproducible FET behavior
\endnote{Without RIE cleaning we observe a large spread in
mobilities, which are on average much lower than the mobility of
FETs made on RIE cleaned substrates. Additionally, we find much
larger hysteresis in the electrical measurements and a large
negative threshold voltage.}.

Freshly grown tetracene crystals placed on RIE-cleaned SiO$_2$
strongly adhere to the substrate. Adhesion only occurs for very
thin crystals ($\sim 1 \ \mu m$) that are sufficiently flexible
and it is probably due to electrostatic forces. The crystals
placed onto substrates are then inspected under an optical
microscope using cross-polarizers. This allows us to select
single-crystalline samples without visible defects for in depth
electrical characterization \cite{Vrijmoeth98}. The top view of a
device fabricated following this procedure is shown in fig.
\ref{FETpicture}c.

We have characterized more than ten single-crystal FETs exhibiting
similar overall behavior. Electrical characterization is performed
in the vacuum chamber of a flow cryostat at a pressure of $10^{-7}
\ mbar$, using a HP4156A Semiconductor Parameter Analyzer. The
measurements shown in this paper have been performed in a
two-terminal configuration.

Fig. \ref{FETcharacteristics} shows the outcome of the
measurements for one of the transistors with highest mobility. The
current increases with increasing negative gate voltage. This
indicates field-effect induced hole conduction, which is the
expected behavior for tetracene. The field-effect mobility is
evaluated in the linear regime of operation, where $I_{sd}$ is
proportional to $V_d$:
\begin{equation}
I_{sd} = \frac{W}{L} \cdot \mu \cdot C_d \cdot (V_g - V_t) \cdot
V_d
 \label{FETmobility}
\end{equation}
Here $C_{d}$ is the capacitance per unit area of the SiO$_2$ layer
and $V_{t}$ is the threshold voltage. From equation
\ref{FETmobility} we obtain the mobility by calculating the
derivative of $I_{sd}$ with respect to both $V_d$ and $V_g$ and
neglecting the dependence of $\mu$ on $V_d$ and $V_g$.

For all the FETs investigated we found room-temperature values of
the mobility larger than $0.01 \ cm^2/Vs$. In many cases $\mu
> 0.1 \ cm^2/Vs$ and the maximum mobility achieved so far is $\mu = 0.4 \ cm^2/Vs$ (fig.
\ref{FETcharacteristics}). This value is better than the highest
mobility recently reported in tetracene TFTs and indicates the
high quality of our single-crystal FETs \cite{Gundlach02}.

The threshold voltage (defined by equation 1) is positive in all
our devices. It ranges from $0 \ V$ to $30 \ V$ and is typically
$V_t \simeq 10 \ V$ \endnote{For holes, a positive threshold
voltage implies conduction through the device in the absence of an
applied gate voltage.}. We do not normally observe a positive
threshold voltage in FETs fabricated without the RIE cleaning, nor
do we observe linear conduction through single crystals contacted
with evaporated gold contacts. These observations suggest that the
current flowing at zero gate voltage is not due to conduction
through the crystal bulk (i.e. crystal doping) but rather to
charge accumulated at the surface \cite{Lin97}. We believe that
the charge accumulation is induced by the electrostatic adhesion
of tetracene crystals to the RIE cleaned SiO$_2$ surface.

As an additional characterization of the single crystals FETs, the
inset of fig. \ref{FETcharacteristics} shows $\log (I_{sd})$
versus $V_g$ at fixed drain voltage ($V_d = -10 \ V$). These data
were measured on a FET with a relatively low mobility of $\mu =
0.05 \ cm^2/Vs$, i.e. they can be considered as typical. From this
measurement we find the subthreshold slope to be $1.6 \
V/\mathrm{decade}$. Normalizing this value to the capacitance of
the dielectric gives $28 \ V \cdot nF / \mathrm{decade} \cdot
cm^2$. These values are comparable to what is found for the best
pentacene TFTs ($15-80 \ V \cdot nF /\mathrm{decade} \cdot cm^2$
\cite{Podzorov03a,Lin97,Gundlach02}).

\begin{figure}[t]
\centering
\includegraphics[width=8.5cm]{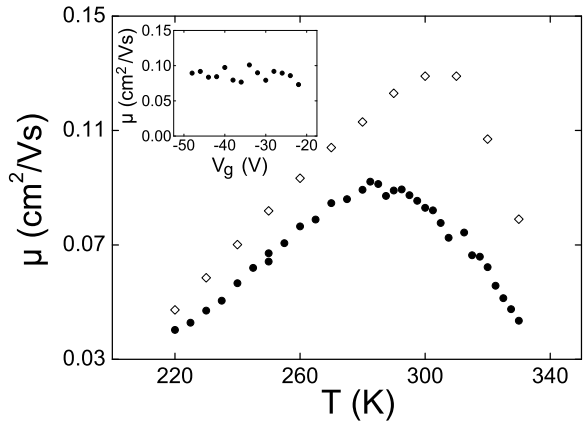}
\caption{Temperature dependence of the field-effect mobility for
two different devices, measured at large negative gate voltage.
The inset illustrates that at large negative gate voltage, $-20$
to $-50 \ V$, where the highest mobility is observed, $\mu$ is
essentially independent of $V_g$. \label{FETmuvstemp}}
\end{figure}

Further proof of the high quality of the tetracene single-crystal
FET is provided by the temperature dependence of the mobility.
Temperature dependent measurements were performed in the range
$220-330 \ K$, since for $T > 330 \ K$ tetracene crystals rapidly
sublime at the pressure present in the measurement chamber, and
for $T < 200 \ K$, a structural phase transition
\cite{Sondermann85} often results in a lowered mobility and in
damage to the devices. Fig. \ref{FETmuvstemp} shows data from two
different FETs. The mobility initially increases with lowering
temperature from $330 \ K$ to $280-300 \ K$, and then decreases
when the temperature is lowered further. Such a non-monotonous
temperature dependence is not usually observed in organic TFTs,
which typically exhibit a thermally activated decrease of $\mu$
with decreasing $T$, over the entire temperature range
investigated \cite{Gundlach02, Vissenberg98}.

The observed temperature dependence of the mobility is
qualitatively similar to the one expected for high-purity organic
crystals in the presence of shallow traps, which is given by $\mu
\propto T^{-n}\exp(-E_t/kT)$ \cite{Silinsh94,Karl99}. As long as
$E_t$ is not much larger than $kT$, this formula accounts for a
non-monotonous temperature dependence of $\mu$ in the temperature
range investigated. The observation of a maximum mobility at room
temperature indicates that $E_t \approx 50-100 \ meV$ (i.e. a few
times $kT$ at room temperature). This is consistent with the weak
gate voltage dependence of the mobility, see the inset of fig.
\ref{FETmuvstemp}, which also points to weak trapping
\cite{Dimitrakopoulos02}.

To obtain more information about the quality of the tetracene
crystals and about the FET fabrication process, we have performed
time-of-flight (TOF) measurements on several thick ($100-200 \ \mu
m$) tetracene crystals grown by the same technique \cite{we}. For
all crystals investigated, the room temperature mobility found in
TOF experiments ranges from $0.5$ to $0.8 \ cm^2/Vs$, comparable
or slightly higher than that obtained from our best FETs. Also the
temperature dependence of the mobility is similar to that observed
in FET measurements \cite{Berrehar76}. Finding comparable values
for the bulk and the surface mobility \endnote{TOF experiments
give a measure of the \textit{bulk} mobility of charge carriers in
the c-direction of the crystal, whereas FET experiments probe the
mobility of charge carriers at the crystal \textit{surface}.}
suggests that the FET fabrication process does not severely
degrade the quality of the crystal surface. We conclude that,
presently, the limiting factor for the mobility of the tetracene
single-crystal FETs is the quality of the as grown crystals.

In conclusion, we have shown that high-quality organic
single-crystal FETs can be realized, whose performance is
comparable to the best existing organic thin film transistors. As
research on organic thin film FETs has now been going on for many
years whereas work on single crystals has just started, we
consider this result to be particularly promising for future
developments.

We are grateful to N. Karl, J. Niemax, and J. Pflaum at the
University of Stuttgart for TOF measurements. This work is
supported by the Stichting FOM. The work of A.F. Morpurgo is part
of the NWO Vernieuwingsimpuls 2000 program.

\end{document}